\documentclass[useAMS,usenatbib]{mn2e}
\input amssym.tex
\usepackage{times}
\usepackage{amstext}
\usepackage{bm}
\usepackage{epsf}
\usepackage{url}
\usepackage{graphics}

\def\hMpc{{\ }h^{-1}{\,}{\rm Mpc}}
\def\nshMpc{h^{-1}{\,}{\rm Mpc}}
\def\LCDM{{\Lambda {\rm CDM}}}

\def\zobov{{\scshape zobov}}
\def\voboz{{\scshape voboz}}
\def\wvf{{\scshape wvf}}
\def\vtfe{{\scshape vtfe}}
\def\dtfe{{\scshape dtfe}}

\newcommand{\twodfigs}{
  \begin{figure*}
    \begin{minipage}{175mm}
      \begin{center}
	\leavevmode
	\epsfxsize=\columnwidth   
      \epsfbox{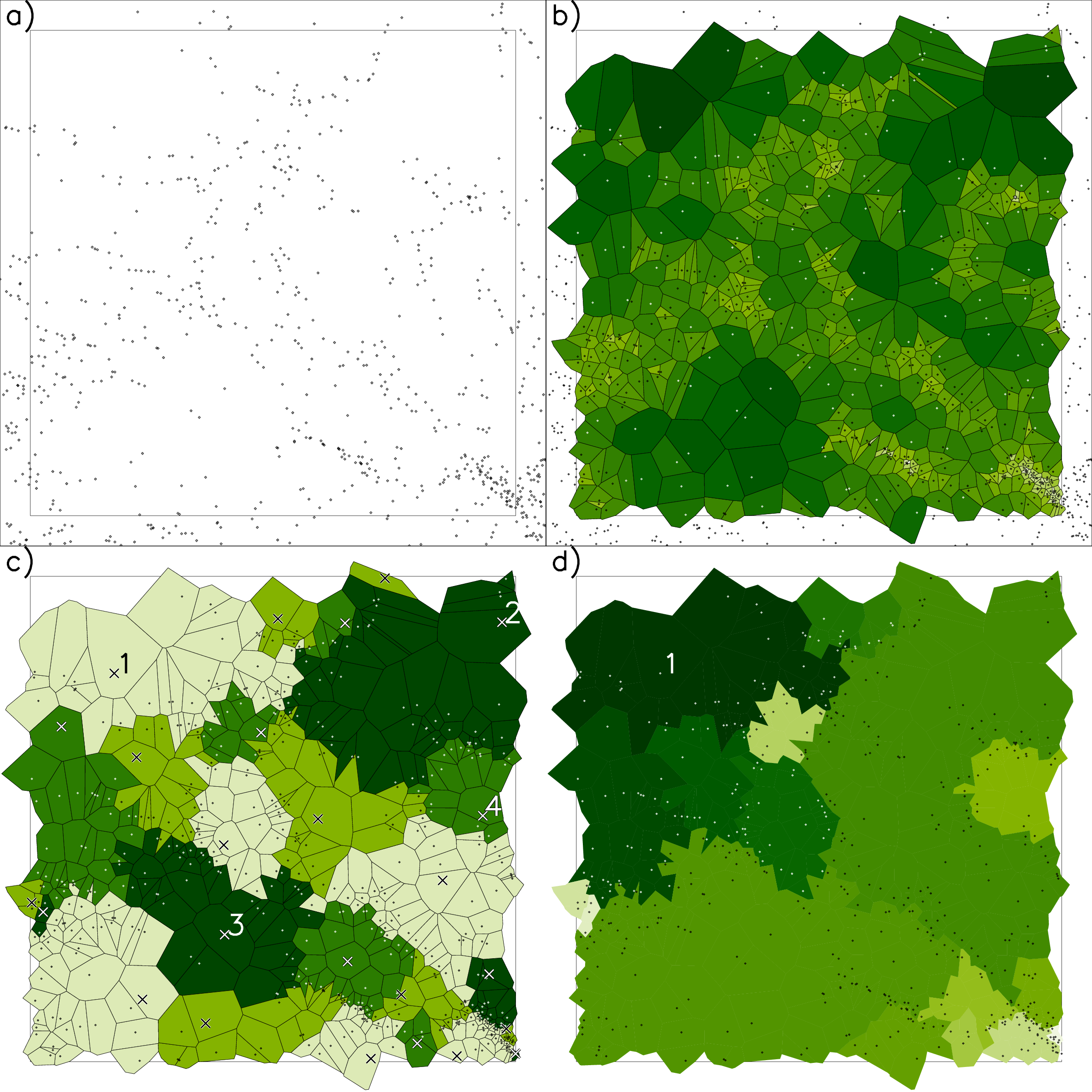}
    \end{center}
    \caption[1]{ \small {\bf(a)} Galaxies \citep[][down to
        $B=-10$]{croton} from a $40\times 40\times 5\ (\nshMpc)^3$
        slice of the AAVFCP region.  The outer boundary is
        $45\hMpc$-square.  The slice is the same size as the
        dark-matter illustration in Fig.\ \ref{zobovoids}, but is at
        an edge of the central 40-$\nshMpc$ cube, not at the centre.
        It was chosen because the voids in this figure are less
        well-defined, and thus richer in structure. {\bf (b)} The 2D
        Voronoi tessellation of galaxies in this slice, with each
        particle's Voronoi cell shaded according to its area.  The
        galaxies outside the inner ($40\hMpc$) boundary are shown
        because they contribute to the tessellation. {\bf (c)} Zones
        of galaxies.  The cores (density minima) of each zone are
        shown with crosses, the different colours merely demarcate
        different zones. {\bf (d)} The growth of void 1, the deepest
        void in the sample.  With analogy to a water tank, the water
        level (density) is increased, and zones the water runs into
        are added to the void.  Colours from dark to light indicate
        the stage at which the zone is added to the void.  The darkest
        colour is the original zone, the next-darkest is the first
        zone or set of zones added, etc.  The only zone that is never
        included is that with the highest-density link to another
        zone, in the lower-right corner.  A measure of the probability
        that each zone-adding event leads to a void that did not arise
        from Poisson noise is shown in Fig.\ \ref{twodprobplot}.
      \label{twodfigs}
    }
  \end{minipage}
  \end{figure*}
}
\newcommand{\voidprobdist}{
  \begin{figure}
    \begin{center}
      \leavevmode
      \epsfxsize=\columnwidth    
      \epsfbox{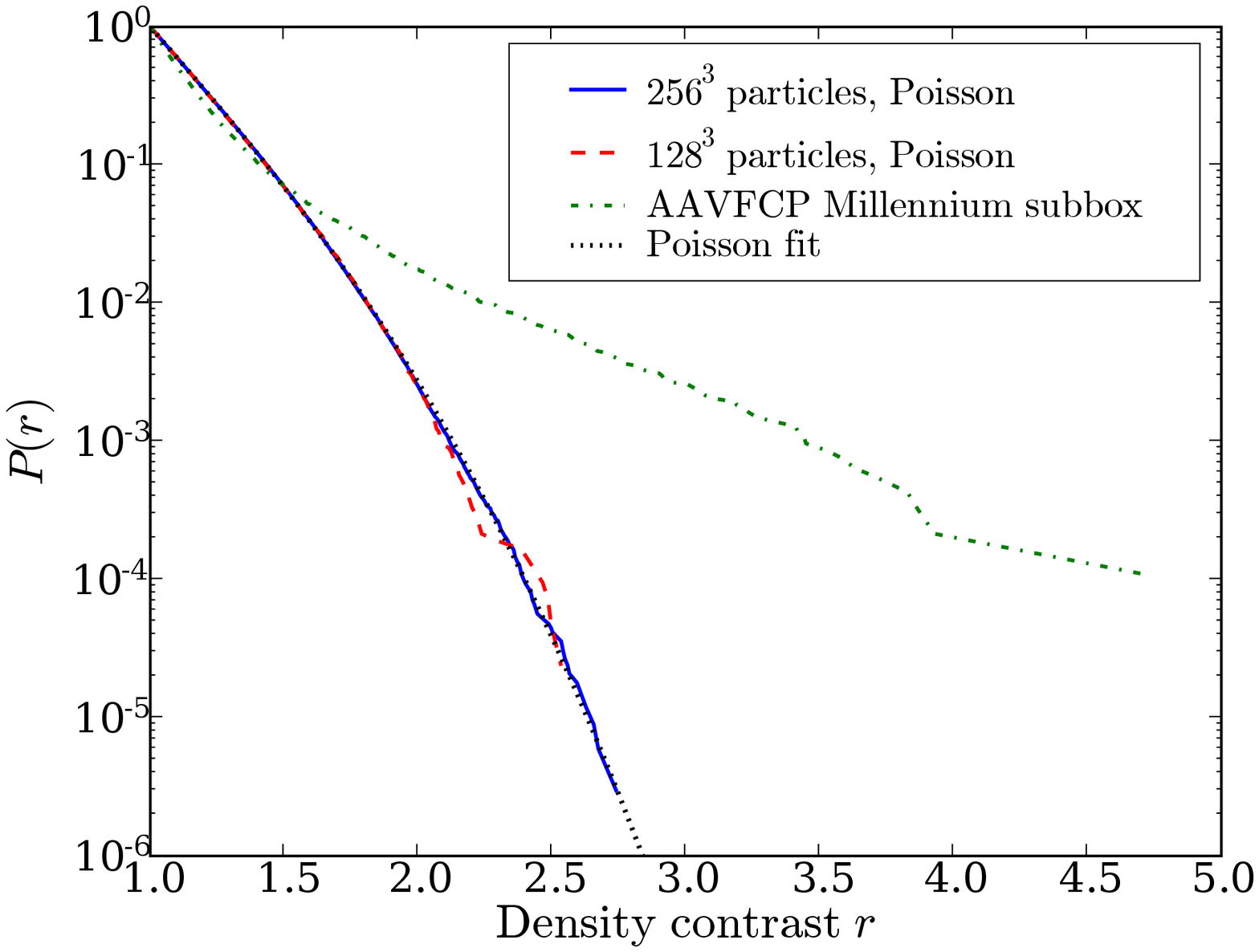}
    \end{center}
    \caption[1]{ \small The cumulative probability function $P(r)$ of
    the ratio $r(v)$ between the lowest density of a zone and
    the density at which water would leak into an adjacent zone that
    is deeper.  The solid and dashed curves show $P(r)$ for
    uniform-density Poisson processes using $256^3$ and $128^3$
    particles.  The curve has the same shape as the number
    of particles increases.  The dotted curve shows the fit in Eq.\
    (\ref{prob}).  The dot-dashed curve shows $P(r)$ for voids
    in the region analysed in the Aspen-Amsterdam Void-Finder
    Comparison Project (AAVFCP).
    \label{voidprobdist}
    }
  \end{figure}
}

\newcommand{\truncvoids}{
  \begin{figure}
    \begin{center}
      \leavevmode
      \epsfxsize=\columnwidth    
      \epsfbox{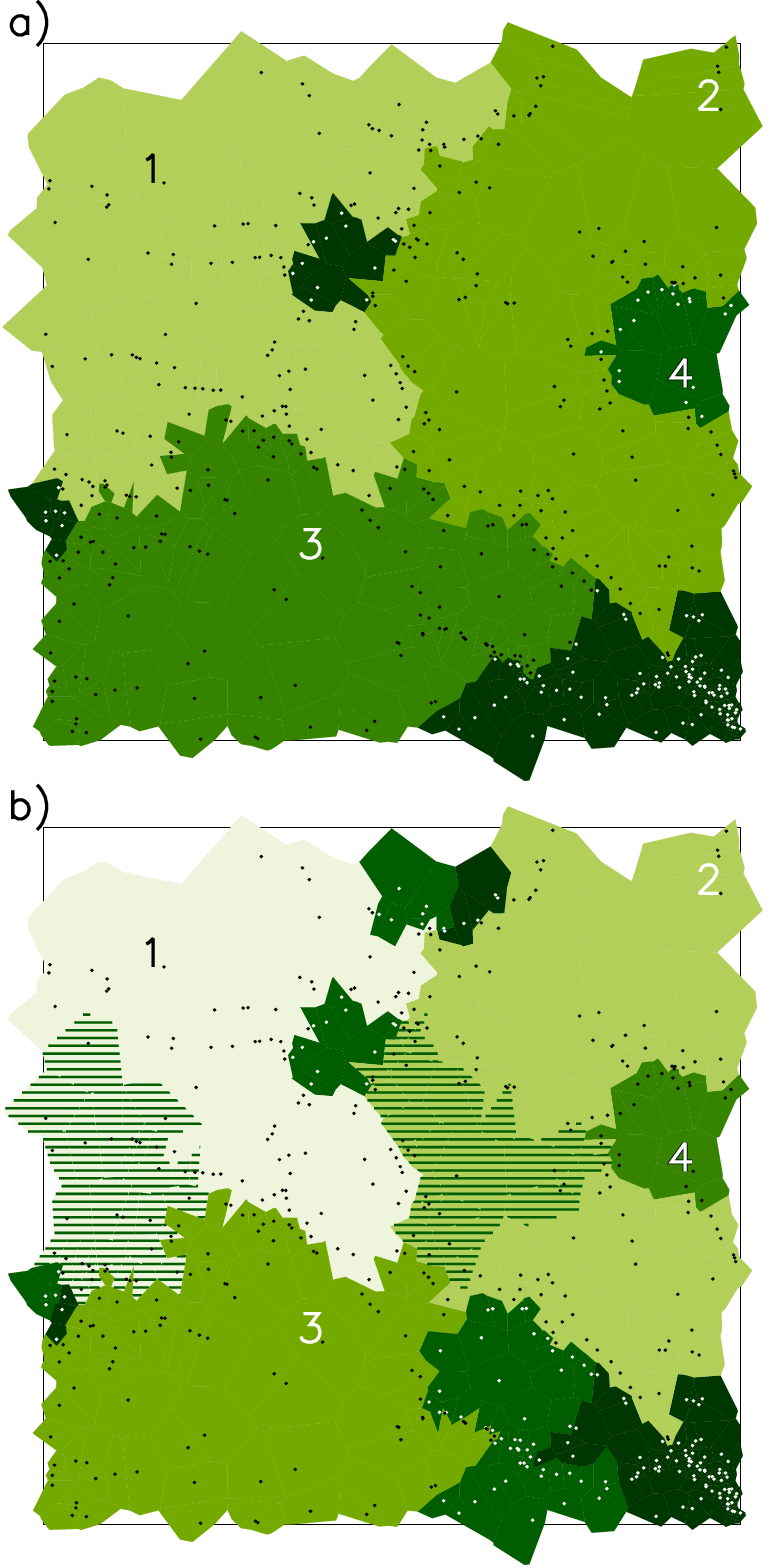}
    \end{center}
    \caption[1]{ \small Two strategies for detecting the edges of
      highly significant voids, as applied to the particle set in
      Fig.\ \ref{twodfigs}.  In {\bf (a)}, discussed in Section
      \ref{siglevel}, the user chooses a significance level at which
      to accept a void (here, 2-$\sigma$).  Voids exceeding this
      threshold stop growing when they encounter another void
      exceeding this threshold.  Particles in the darkest regions
      belong to no void over 2 $\sigma$. In {\bf (b)}, discussed in
      Section \ref{mostsig}, a most-probable extent is found for each
      void.  Zones are coloured according to their significance.
      Zones in the deepest, 4-$\sigma$ void are lightest; zones
      included in 3-, 2-, and 1-$\sigma$ voids are coloured
      increasingly darkly.  Particles in the darkest region belong to
      no void over 1 $\sigma$.  The hatched regions are 1-$\sigma$
      subvoids.
      \label{truncvoids}
    }
  \end{figure}
}
\newcommand{\twodprobplot}{
  \begin{figure}
    \begin{center}
      \leavevmode
      \epsfxsize=\columnwidth    
      \epsfbox{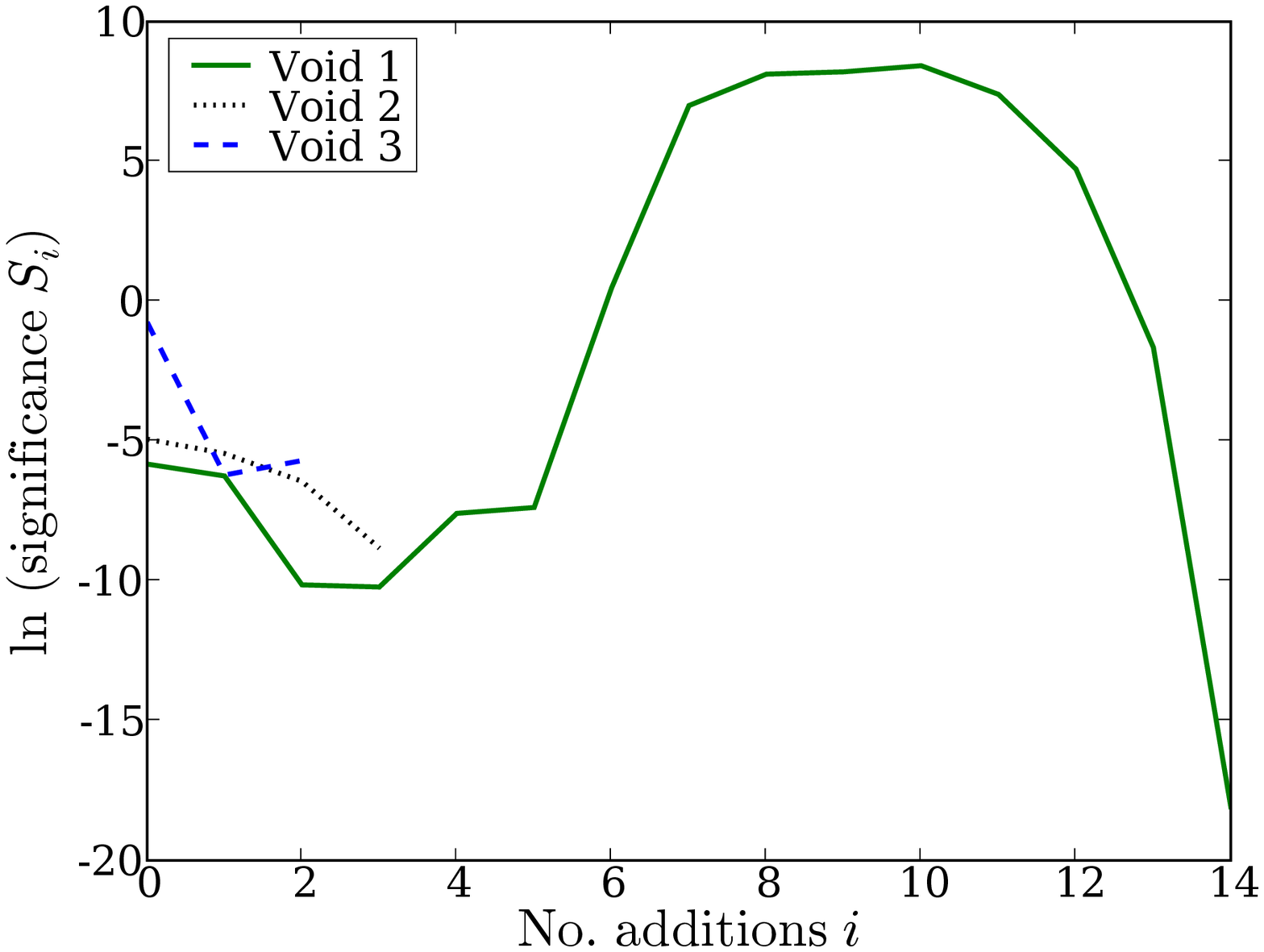}
    \end{center}
    \caption[1]{ \small Significances of various extents of the three
      voids in Fig.\ \ref{twodfigs} that encompass more than their
      central zone.  The extents at the minima of these curves are
      shown in Fig.\ \ref{truncvoids}b, with the exception of void 1.
      For void 1, the extent shown in Fig.\ \ref{truncvoids}b has a
      number of additions $i=3$, the minimum of the curve excluding
      the high-$i$ ramp into high-density regions.
      \label{twodprobplot}
    }
  \end{figure}
}

\newcommand{\aavfcp}{
  \begin{figure}
    \begin{center}
      \leavevmode
      \epsfxsize=\columnwidth    
      \epsfbox{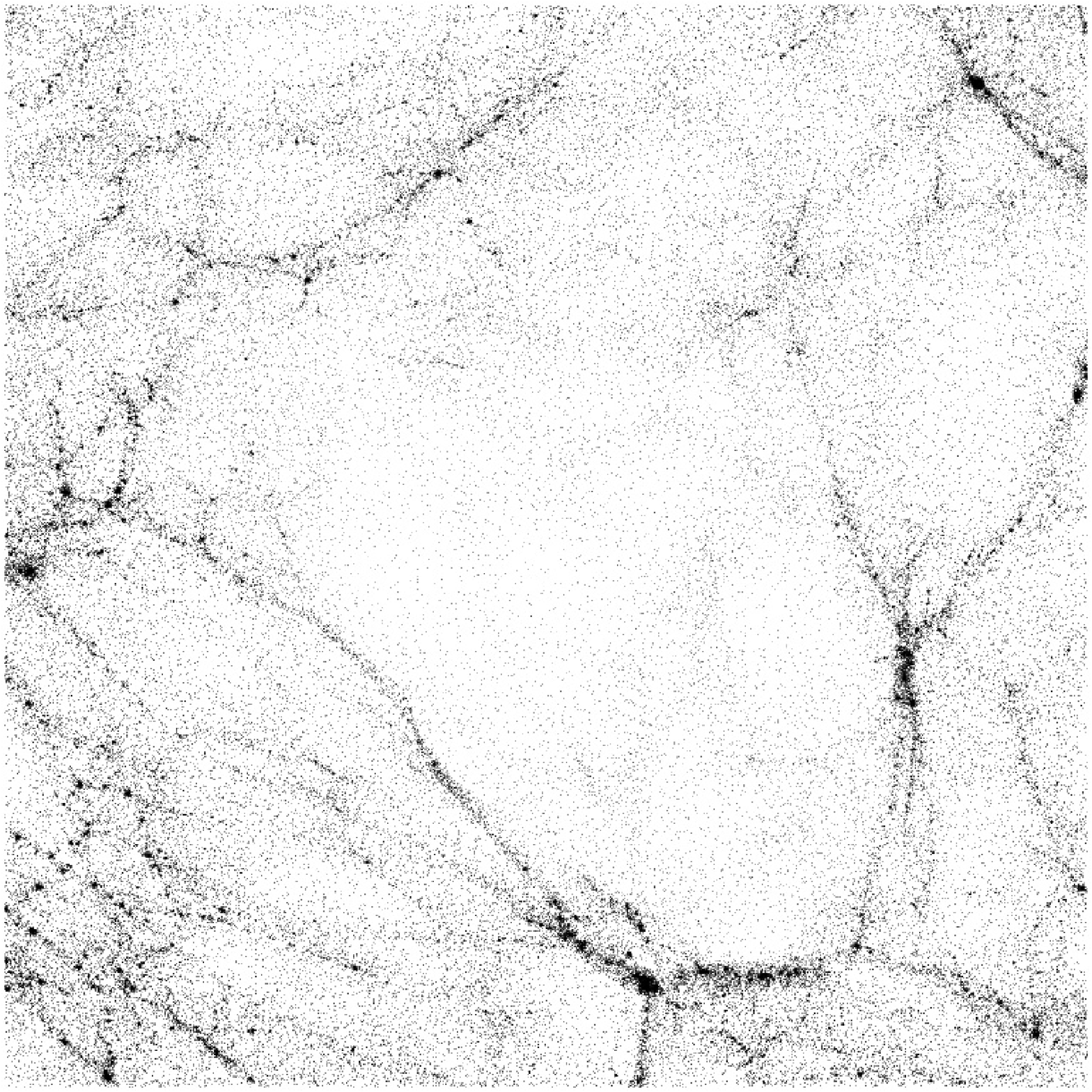}
    \end{center}
    \caption[1]{ \small A 5-$\nshMpc$-wide slice through the the inner
    40-$\hMpc$ cube analysed for the AAVFCP.  This and the following
    figure were produced using Nick Gnedin's {\scshape ifrit}
    software, available at \url{http://home.fnal.gov/~gnedin/IFRIT/}.
    \label{aavfcp}
    }
  \end{figure}
}
\newcommand{\zobovoids}{
  \begin{figure}
    \begin{center}
      \leavevmode
      \epsfxsize=\columnwidth    
      \epsfbox{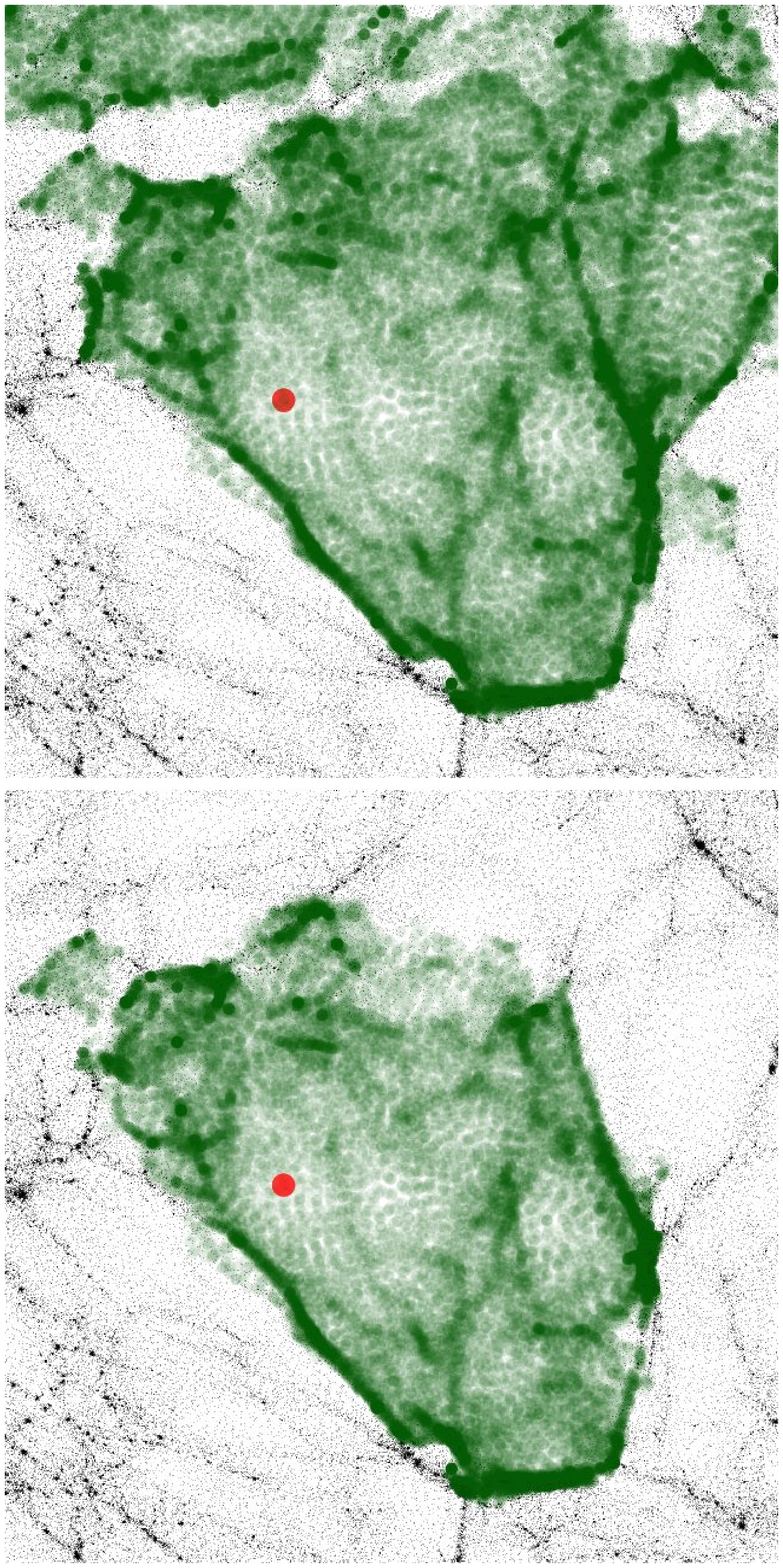}
    \end{center}
    \caption[1]{ \small The largest, and most significant, void
    \voboz\ found the inner 40-$\hMpc$ AAVFCP cube.  Green, diffuse
    particles are in the void; black particles are not.  The large red
    dot is the core (minimum-density) particle in the void.  The top
    panel shows the full void. The bottom panel shows the void as
    truncated as described in Section \ref{siglevel}, using a
    significance level of 5-$\sigma$; it is the same as the
    most-probable extent of the void as described in Section
    \ref{mostsig}, if the `steeper fit' in Fig.\ \ref{probplot}, Eq.\
    (\ref{probfudged}), is used.
    \label{zobovoids}
    }
  \end{figure}
}

\newcommand{\probplot}{
  \begin{figure}
    \begin{center}
      \leavevmode
      \epsfxsize=\columnwidth    
      \epsfbox{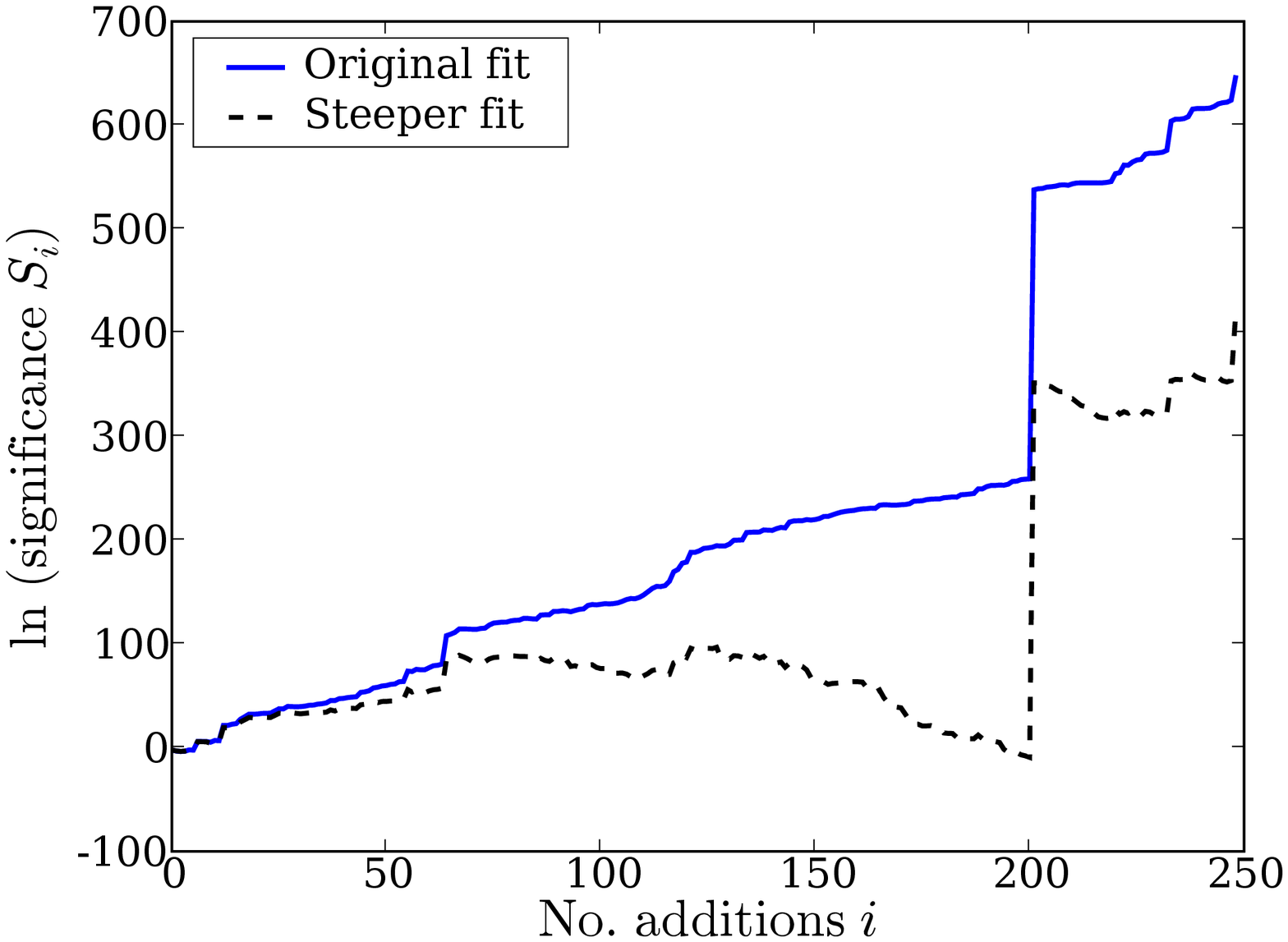}
    \end{center}
    \caption[1]{ \small Significances of possible void extents for the
      void shown in Fig.\ \ref{zobovoids}.  The most-probable extent
      (a minimum on the curve) is at $i=2$ using the original fit to
      the void probability function, Eq. (\ref{prob}).  To achieve a
      minimum at $i=200$ (the extent shown in the bottom panel of
      Fig.\ \ref{zobovoids}), a steeper fit such as Eq.\
      (\ref{probfudged}) must be used.  The simulations used for Fig.\
      \ref{voidprobdist} are too small to probe the abundance in
      Poisson simulations of \zobov\ voids with as high density
      contrast as this void.
      \label{probplot}
    }
  \end{figure}
}
\newcommand{\density}{
  \begin{figure}
    \begin{center}
      \leavevmode
      \epsfxsize=\columnwidth    
      \epsfbox{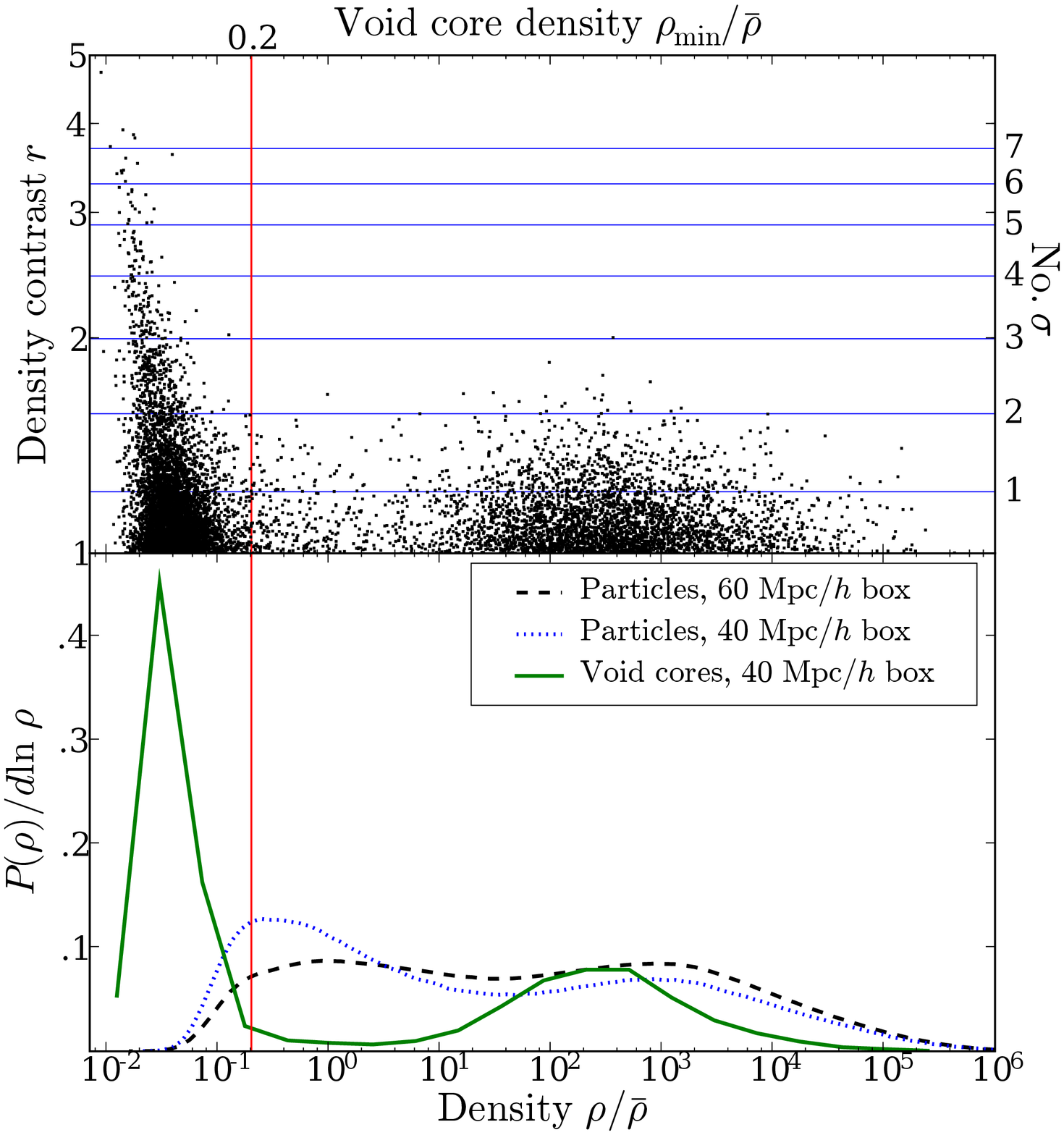}
    \end{center}
    \caption[1]{ \small {\it Top}. For \zobov\ voids in the
      40-$\nshMpc$ AAVFCP region, a scatter plot of the minimum
      density $\rho_{\rm min}$, and the density contrast $r$ (the
      ratio of $\rho_l$, the density at which a void would merge with
      a deeper void, and $\rho_{\rm min}$).  This plot shows two
      populations, one that satisfies the `physical' significance
      criterion, $\rho_{\rm min}/\bar{\rho} < 0.2$, and another that
      does not.  The high-density population contains only one void
      above $3\sigma$, whereas $10$ would be expected from 3765
      Poisson voids.

      {\it Bottom}.  Probability density functions (PDF's) of particle
      densities in the AAVFCP region.  The dashed black curve shows
      the PDF from particles in the full 60-$\nshMpc$ box; the dotted
      blue curve uses only particles in the inner 40-$\nshMpc$ box.
      The inner box has fewer haloes per unit volume; this explains
      the lower high-density peak, and higher low-density peak, in the
      inner box.  The solid green curve shows the PDF of void
      minimum-density particles.  Peaks in the PDF using the full
      particle set (e.g.\ the dashed blue curve) seem to give sharper
      peaks in the PDF of void core-particle densities (the green
      curve) at slightly smaller densities.
    \label{density}
    }
  \end{figure}
}

\begin{document}
\title[{\scshape zobov}: a parameter-free void finder] {\zobov: a parameter-free void-finding algorithm}
\author[Mark C.\ Neyrinck]
{Mark C.\ Neyrinck$^1$\\
$^{1}$Institute for Astronomy, University of Hawaii, Honolulu, HI 96822, USA\\
  email: {\tt neyrinck@ifa.hawaii.edu}}
\pubyear{2008}

\bibliographystyle{mnras}
\maketitle
  
\begin{abstract}
  \zobov\ (ZOnes Bordering On Voidness) is an algorithm that finds
  density depressions in a set of points, without any free parameters,
  or assumptions about shape.  It uses the Voronoi tessellation to
  estimate densities, which it uses to find both voids and subvoids.
  It also measures probabilities that each void or subvoid arises from
  Poisson fluctuations.  This paper describes the \zobov\ algorithm,
  and the results from its application to the dark-matter particles in
  a region of the Millennium Simulation.  Additionally, the paper
  points out an interesting high-density peak in the probability
  distribution of dark-matter particle densities.

\end{abstract}
\begin {keywords}
  large-scale structure of Universe -- methods: data analysis -- cosmology: theory
\end {keywords}

\section{Introduction}

Voids are an essential component of the cosmic web \citep{bkp} of
matter in the Universe on several-Megaparsec scales.  They are
fascinating from an information-theoretic viewpoint, as a probable
component of efficient descriptions of large-scale structure in the
non-linear regime.  Voids also provide useful tools for studying
cosmology and galaxy formation.  The way in which matter inside a void
flows away from its centre holds information about cosmological
parameters such as the matter density $\Omega_m$ and dark-energy
density $\Omega_\Lambda$ \citep{dekelrees,bvdw,flichetriay}, and about
the clustering, if it exists, of dark energy \citep{mss}.  Also,
measuring the evolution of void ellipticities can give constraints on
the dark-energy equation of state \citep{leepark07}.  The existence of
large voids has been invoked to explain the `cold spot' on the Cosmic
Microwave Background (CMB) \citep{rbw}; anomalously low large-angle
CMB anisotropies \citep{is}; and even the apparent accelerating
expansion of the Universe \citep[e.g.][]{moffat,cel,abnv}.  Voids are
also relatively pristine laboratories to study galaxy formation and
evolution, containing the most isolated galaxies in the Universe.  For
example, \citet{peebles} has pointed out that there seem to be fewer
galaxies in voids than cosmological simulations predict.  Even if this
is not a discrepancy with the underlying $\LCDM$ cosmology, it
contains valuable information about galaxy formation.

Despite these useful features of voids, they are not currently in the
forefront of cosmological probes.  One reason for this is that there
remains no standard definition of them.  The Aspen-Amsterdam
Void-Finder Comparison Project \citep[AAVFCP]{aavfcp} makes an
important first step in exploring differences and similarities in void
definitions, but still, a consensus about how to define a void does
not exist.  Here I present \zobov\ (ZOnes Bordering On Voidness), a
void finder whose features, I believe, are appealing enough that it
represents a net contribution toward that consensus, rather than
simply adding another alternative to reconcile with the others.

Many void finders define voids as spheres, or unions of a finite
number of spheres or other shapes \citep[e.g.][]{kf,maet,hv,colb}.
This definition has some theoretical justification, since underdense
regions expanding in a homogeneous background tend to become more
spherical with time \citep{icke}.  Also, this definition is
geometrically simple.  However, the real Universe consists of many
underdense regions that collide with each other and produce voids that
are often more polyhedral than spherical \citep[e.g.][]{ivdw}, or even
more generally shaped \citep{sfhk}. \zobov\ imposes no prejudice about
the shape, or even topology, of a void.  Some other void finders
\citep*[e.g.][]{ep,am,pb,sfhk,hahn,ac,pwj} also define voids with no,
or only weak, rules about their shapes.

\zobov\ aims to find voids from a set of points with as few
restrictions as possible.  Conceptually, a \zobov\ void is simply a
density minimum with a depression around it.  \zobov\ has no free
parameters.  However, the ambiguity in void finding, when applied to
noisy data, must be placed somewhere.  With no free parameters to
tune, \zobov\ returns many (indeed, mostly) shallow, hardly visible
voids.  However, \zobov\ measures a statistical significance for each
void.  A physical significance criterion can also be used, requiring
that a void's minimum density $\rho_{\rm min}/\bar{\rho} < 0.2$.  This
is a characteristic density of a void in an Einstein-de Sitter model
(with only a slightly different density in $\LCDM$), obtained using a
top-hat spherical expansion model \citep{bdglp,svdw}.  For the
abstract problem of finding voids in a particle distribution, this
strategy of returning all possible voids, even statistically dubious
ones (as long as they are marked as such), seems more satisfying than
making an arbitrary choice of free parameters.  For many applications,
though, actual use of a \zobov\ void catalogue might require another
arbitrary choice, about the level of significance at which to accept a
void.  Another philosophical difference between \zobov\ and most void
finders is that \zobov\ returns subvoids along with voids.

\zobov\ is an inversion of an `almost-parameter-free' dark-matter-halo
finder, called \voboz.  \citep[VOronoi Bound Zones;][NGH]{voboz}.  The
major change in \zobov\ is that it looks for density minima instead of
maxima.  In fact, the \voboz-\zobov\ algorithm is perhaps
better-suited for void finding than halo finding.  This is because the
algorithm typically detects highly nonspherical shapes; to get roughly
spherical, virialized haloes, \voboz\ trims their edges with a
boundedness criterion using particle velocities.  It is in this step
that \voboz\ loses its pure parameter-freedom.

Perhaps the existing void finder most similar to \zobov\ is \wvf\
\citep[the Watershed Void Finder;][]{pwj}.  Both use tessellation
techniques to measure densities, and both use the `watershed' concept,
defining voids with analogy to catchment basins in a density field.
However, \wvf\ uses several clever techniques from the field of
mathematical morphology to smooth the particle density before defining
voids, while \zobov\ analyses the raw, unsmoothed data.  There are
other methods that use tessellation techniques to find clusters
\citep{rbfn, barkhouse, soechting, vb, mev} or voids \citep{g05,ac}.

First I will discuss the \zobov\ algorithm, and then I will discuss
its application to dark-matter particles in a region of the Millennium
simulation \citep{mill}.  Some of these results appear in AAVFCP,
where they are also compared to the results of other void finders.
Finally, I will discuss what I feel are the unique strengths and
weaknesses of \zobov, and what could be done to improve it.

\section{Method}
\label{method}

The \zobov\ algorithm is the same as the first two steps of the
\voboz\ (NGH) algorithm, except that it searches for density minima
instead of maxima.  

\subsection{Particle density and adjacency measurement}
The first step is the density estimation at each dark-matter particle,
using what \citet{schaap} calls the Voronoi Tessellation Field
Estimator (\vtfe).  Tessellation methods for density estimation are
widely used in many fields \citep[e.g.][]{brown,ord,boshh}.  A good
reference on this topic is provided by \citet{okabe}; see \citet{vdws}
for a review specific to large-scale structure.  The \vtfe\ (along
with its dual, the \dtfe) gives arguably the most local possible
density estimate that has meaningful information.  The Voronoi
tessellation divides space into cells around each particle, with the
cell around particle $i$ defined as the region of space closer to
particle $i$ than to any other particle.  The density estimate at
particle $i$ is $1/V(i)$, where $V(i)$ is the volume of the Voronoi
cell around particle $i$.  The Voronoi tessellation also gives a
natural set of neighbours for each particle (the set of particles
whose cells neighbour $i$'s cell), which \zobov\ uses in the next
step.

Figure \ref{twodfigs}a shows a set of particles in 2D, corresponding
to galaxies in a slice of the Millennium simulation.  Figure
\ref{twodfigs}b depicts the Voronoi tessellation of this set of
particles, with the Voronoi cells shaded according to area.

\twodfigs
\subsection{Zoning}
The second step in \zobov\ is the partition of the set of particles
into zones around each density minimum.  This is done partly for
computational speed, and partly to compress the information in the
dataset.  A minimum is a particle with lower density than any of its
Voronoi neighbours.  \zobov\ sends each particle to its lowest-density
neighbour, repeating the process until it arrives at a minimum.  A
minimum's {\it zone} is the set of particles which flow downward into
it, and a zone's {\it core} is the minimum-density particle of the
zone.  Figure \ref{twodfigs}c shows how \zobov\ partitions the
particles in the previous panels into zones.  These zones could
conceivably be called voids.  Because of discreteness noise, though,
many zones are spurious, and others are only the central parts of what
are picked out as voids by eye.  Thus, it is necessary to join some
zones together to form the final voids.

\subsection{From zones to voids}
\label{zonestovoids}
Zones are joined as follows.  Imagine a 2D density field (represented
as height) in a water tank; see, for example, Fig.\ 1 of \citet{pwj}.
For each zone $z$, the water level is set to $z$'s minimum density,
and then raised gradually.  Water may flow, along lines joining
Voronoi neighbours, into adjacent zones, adding them to the {\it void}
defined around the zone $z$.  The process stops when water flows into
a deeper zone (with a lower minimum than $z$'s), or if $z$ is the
deepest void, when water floods the whole field.  The final void
corresponding to $z$ is defined as the set of zones containing water
just before this happens.  

The minimum-density (core) particle of the original zone is also the
minimum-density particle of the zone's void.  Many low-significance
zones fail to annex surrounding zones as they attempt to grow; a zone
in this situation has a void equal to itself.  The density (water
level) at which water flows into a deeper zone is recorded as
$\rho_l(z)$ ($l$ stands for `link' to a deeper zone).

Figure \ref{twodfigs}d shows the stages of growth that the deepest
void in the set of particles undergoes.  Successively lighter colours
shade zones added when the density level reaches successively higher
levels.  Since this is the deepest void, its last extent encompasses
the whole simulation, except for the zone with the highest-density
particle separating it from other zones, in the lower-right corner of
the figure.  This does not mean that other voids are not detected;
they are subvoids of this largest void.  Still, to form voids
conforming better to intuition, a further criterion could be used to
halt the growth of voids containing several zones.  I will come back
to this issue after discussing the statistical significance of voids,
which will be useful for defining their edges.

This way of defining voids can lead to surprising void topologies and
shapes.  For example, if a set of particles consists of a clump
surrounded by a low, uniform-density background, everything but the
clump will be detected as a void.  Also, even a single low-density
particle along a wall between two visually apparent voids might cause
\zobov\ not to detect them separately, but instead to detect a single,
dumbbell-shaped void.  However, many (about 16) particles directly
participate in each particle's density estimate.  Thus, such a hole in
the wall between voids would have to be a conspiracy of many
particles, and would likely look like a significant hole by eye, as
well.  \zobov\ operates under an implicit assumption that the
discreteness noise is similar to that in a Poisson density-sampling,
and \zobov\ could give surprising results if particles are carefully
arranged to fool it.

\subsection{Statistical significance of voids}
\label{statsig}

The probability that a void $v$ is real is judged according to its
density contrast, i.e.\ the ratio $r(v)$ of $\rho_l(v)$, the
minimum-density particle on a ridge beyond which is a deeper void, to
$v$'s minimum density, $\rho_{\rm min}$.  This is not the only
conceivable way to judge the significance of a void.  But it is
simple, and the probabilities it returns roughly align with what
visual inspection would suggest.

\voidprobdist

The density contrast $r$ is converted to a probability by comparing to
a Poisson particle distribution.  Several statistical properties of
Voronoi diagrams applied to Poisson-sampled uniform density
distributions are well-understood.  For example, the distribution of
Voronoi cell volumes is well-approximated by a gamma distribution
\citep{kiang}, and the average number of Voronoi neighbours
($48\pi^2/35+2\approx 15.54$), is even known analytically
\citep{okabe}.  Unfortunately, the distribution of contrasts of
density depressions in a Poisson Voronoi diagram is not known.  It
seems difficult to model analytically or from known results, since
each depression has an unknown number of particles, whose estimated
densities depend on each other in a complicated way.  Therefore, for
\zobov, this distribution is measured in a Monte-Carlo fashion from
Poisson sampling.

Let the cumulative probability $P(r)$ be the fraction of voids in a
Poisson particle distribution with density contrast greater than $r$.
Figure \ref{voidprobdist} shows $P(r)$ as a function of $r$ for two
cubic Poisson simulations (assuming periodic boundary conditions),
with $128^3$ and $256^3$ particles.  It also shows the following fit
to $P(r)$:
\begin{equation}
P(r) = \exp[-5.12(r-1)-0.8(r-1)^{2.8}].
\label{prob}
\end{equation}
This $P(r)$ gives an estimate of the likelihood that a void with
density contrast $r$ could arise from Poisson noise, i.e.\ that it is
fake.  Table \ref{ratios} shows density contrasts corresponding to the
first seven `sigmas' (with analogy to a Gaussian distribution),
calculated using this fit.  The fit may be trusted to roughly $r = 3$,
beyond which there is no Poisson data.  In NGH, we found that the
analogous significance measure for haloes seems to lose its meaning at
some point between the 4- and 7-$\sigma$ level anyway.  That is,
7-$\sigma$ haloes are not visibly more robust than 4-$\sigma$ ones.

\begin{table}
  \center
  \begin{tabular}{l|l|l|l|l}
    \hline
    $\sigma$ & $P(r)$ & $r$ & Voids & Voids $(\rho_{\rm min} < 0.2)$\\
    \hline
    0 & 1 & 1 & 9308 & 5543\\
    1 & 0.317 & 1.22 &  2362 & 1722\\
    2 & $4.55 \times 10^{-2}$ & 1.57 &  525 & 502\\
    3 & $2.70 \times 10^{-3}$ & 2.00 &  164 & 163\\ 
    4 & $6.33 \times 10^{-5}$ & 2.45 &  64 & 64\\
    5 & $5.73 \times 10^{-7}$ & 2.89 &  29 & 29\\
    6 & $1.97 \times 10^{-9}$ & 3.3 &  13 & 13\\
    7 & $2.56 \times 10^{-12}$& 3.7 &  5 & 5\\
    \hline
  \end{tabular}

  \caption{ \small Void abundances for various density contrasts $r$
    in a Poisson particle simulation, and in the AAVFCP region.  With
    analogy to a Gaussian distribution, the first two columns list the
    levels of probability corresponding to different $\sigma$'s.  The
    third column ($r$) gives the density contrast with abundance
    $P(r)$ in a Poisson simulation, calculated using Eq.\ \ref{prob}.
    The fourth column gives the number of voids exceeding density
    contrast $r$ in the 40-$\nshMpc$ AAVFCP region.  The last column
    adds the constraint that the minimum density of the void
    $\rho_{\rm min}<0.2$, in units of the mean density.  The last two
    columns are illustrated in the top panel of Fig.\ \ref{density}.
    }
  \label{ratios}
\end{table}

Figure \ref{voidprobdist} also shows the cumulative distribution of
the density contrast $r$ for a region of Millennium simulation,
discussed in Section \ref{results}; results for this region are listed
in Table \ref{ratios}, as well.  Possibly, a natural place to stop
accepting voids as real in this data set would be where $r$ goes below
the curves' intersection, at $r\approx 1.5$.  However, this occurs at
about a 2-$\sigma$ level, which seems quite low.

In the $256^3$-particle Poisson simulation, \zobov\ detected 335025
voids; thus, the average number of particles in a zone is 50.1.  This
also means that one out of every 50.1 particles is a density minimum.
The average number of particles in a void is greater than 50.1,
though, since a void may be comprised of many zones.

The 2D version of Eq.\ (\ref{prob}) is
\begin{equation}
P(r) = \exp[-2.6(r-1)].
\label{prob2d}
\end{equation}
This fit is based on a rather small set of 256$^2$ uniformly
Poisson-distributed particles.  It only extends to $P(r) \approx
10^{-3}$ (about 3 $\sigma$).  This fit gives $r=1.44$, 2.19, 3.3, 4.7,
7, 9, and 11 for significance levels of 1-7$\sigma$.

In addition to the statistical probability criterion, there is a
simple physical criterion to use.  The natural dark-matter density
associated with a top-hat void that has undergone spherical expansion
is $\rho_{\rm void} \approx 0.2$ (hereafter, densities are assumed to
be in units of the mean density) at redshift 0.  Because the densities
of galaxies and dark matter differ in general, this criterion may be
inappropriate to use for galaxies.  This could be incorporated into
the significance measure, for example by multiplying the probability
the void is fake statistically by the probability of getting its
core-particle (minimum) density $\rho_{\rm min}$ in a Poisson Voronoi
diagram with density 0.2.  However, in the AAVFCP sample, all \zobov\
voids statistically significant at the $\gtrsim 3 \sigma$ level have
core densities $\rho_{\rm min} < 0.2$ anyway (see Fig.\
\ref{density}).  Also, the population with $\rho_{\rm min}<0.2$ is
quite distinct; there are few voids close to $\rho_{\rm min} = 0.2$.
So, a simple cut-off at $\rho_{\rm min}=0.2$ may suffice as a physical
criterion, redundant if only large-significance voids are used.

\subsection{Defining the edges of voids}
\label{edgedetect}
As noted above, the deepest \zobov\ void in a set of particles will
encompass all zones except the one with the highest-density ridge
separating it from other zones.  There are (at least) three ways to
deal with this situation.

The first option is to do nothing further.  The raw \zobov\ results
would then consist of a large void, with several subvoids (and
sub-subvoids, etc.) of varying significance levels.  A zone can belong
to multiple voids and subvoids.  This option could be appealing in its
simplicity, and is well-suited to the physical hierarchy that voids
are thought to have in the Universe \citep{dubinski, svdw, fp}.
However, the following two options likely produce more practically
usable sets of voids.

\subsubsection{Specifying a significance level}
\label{siglevel}
The second option is to excise subvoids exceeding a particular
significance level (e.g., 5-$\sigma$) from parent voids.  If a subvoid
is removed from a void, then all zones which join the parent void in
the same accretion event as that subvoid, or in subsequent ones, are
also removed.  This option is a natural choice if disjoint voids are
desired, which is traditionally the case.

Figure \ref{truncvoids}a shows the result of this procedure, applied
to the set of particles in Figure \ref{twodfigs}, using a 2-$\sigma$
threshold of $r=2.19$.  According to Eq.\ \ref{prob2d}, voids 1 and 2
are significant at the 4-$\sigma$ level, and voids 3 and 4 are
significant at the 3- and 2-$\sigma$ levels.  The darkest regions
belong to no void over $2\sigma$.

\subsubsection{Determining the most probable extent of voids}
\label{mostsig}
The third option is to use the density contrasts of voids and subvoids
to define a most-probable extent of voids.  Suppose a zone $z$ has a
sequence of extents, $v_i$.  For example, Fig. \ref{twodfigs}d shows
the various possible extents for void 1 in that particle set.

At each zone-adding event, define a significance $S_i$.  The
significance of zero zone additions $S_0 \equiv P[r(z)]$, the
probability of zone $z$'s density contrast $r(z)$ arising in a Poisson
particle distribution.  Note that the density ratios $r(z) =
\rho_l(z)/\rho_{\rm min}(z)$ and $r(v)= \rho_l(v)/\rho_{\rm min}(z)$
can differ.  $\rho_l(v)$ is the lowest density among ridge particles
linking $z$ to a deeper zone (perhaps with a path through other
zones), while $\rho_l(z)$ is the lowest density among ridge particles
linking $z$ to any of its neighbouring zones.

\truncvoids
\twodprobplot

Call the void after the $i$th zone-adding event $v_i$ ($v_0 \equiv
z$), and call the set of zones to be added in the $(i+1)$st
zone-adding event $Z_{i+1} = \{z_{i+1,j}\}$.  To judge the wisdom of
the $(i+1)$st addition, compare the probability that $v_i$ and all of
the zones in $Z_{i+1}$ are individually fake to the probability that
their union, $v_{i+1} = v_i + Z_{i+1}$, is fake.  Given $S_i$, define
$S_{i+1}$ as
\begin{equation}
  S_{i+1} = S_i\frac{P[r(v_{i+1})]}{P[r(v_i)] \prod_j P[r(z_{i+1,j})]}.
\label{signif}
\end{equation}
Here, $r(v_i)=\rho_l(v_i)/\rho_{\rm min}(z)$, where $\rho_l(v_{i})$ is
the the lowest density among particles on the ridge separating $v_i$
from the set of prospective new zones $Z_{i+1}$.  If the new zone set
$Z_i$ consists of other entire voids (i.e.\ sets of zones) with
subvoids, only the voids, and not the subvoids, enter the product in
the denominator.

For example, zone 3 in Figure \ref{twodfigs}c, and its two
neighbouring zones below and to the left of it (call them $3^\prime$
and $3^{\prime\prime}$), are separated by an insignificant (below
1-$\sigma$) density ridge, undetectable by eye in the original
particle set.  The probability that all three are fake (separately
arose from Poisson noise) is $P[r(3)] P[r(3^\prime)]
P[r(3^{\prime\prime})] = P(1.28)P(1.11)P(1.06) = 0.30$, using Eq.\
(\ref{prob2d}).  The probability that their union is fake is
$P[r(3+3^\prime+3^{\prime\prime})] = P(3.57) = 0.0013$.  Since the
latter is rarer, the union is favoured statistically.  In the $S_i$
notation, these probabilities are normalized differently; to get the
expressions in this paragraph, multiply through by the probabilities
in the denominator of Eq.\ (\ref{signif}).

Figure \ref{twodprobplot} shows the significances of the various
possible extents for the three large voids in Fig.\ \ref{twodfigs}.
It shows the first dip in void 3's likelihood of fakeness when the
first group of zones is added, discussed in the previous paragraph.
The second (and last) prospective addition, the last one before a
deeper zone is encountered, gives an upturn in void 3's curve.  Thus,
the last addition is not favoured statistically.  However, using the
method of Section \ref{siglevel}, this extra zone is included in zone
3.

Void 1, the deepest void, has the longest curve in
Fig.\ \ref{twodprobplot}.  Its probability of fakeness reaches a local
minimum after the third addition of zones.  In accord with intuition,
the curve then increases again, but then as the densest zones in the
figure eventually get included, the density contrast grows sharply,
making the curve plunge.  \zobov\ is detecting everything except the
dense points in the lower-right corner as a highly significant void.
To prevent voids from growing into haloes, a density limit for links
between zones may be set.  For dark matter, a natural value for this
limit would be $\rho_{l,{\rm max}}=0.2$.  Alternatively, one might simply
accept the lowest minimum before the ramp downward at the end.

Figure \ref{truncvoids}b shows most-probable \zobov\ void extents for
the 2D particle set.  Zones are coloured according to their
significance level.  Zones in the deepest, 4-$\sigma$ void (defined
using three zone-addition events) are lightest; zones included in 3-,
2-, and 1-$\sigma$ voids are coloured increasingly darkly.  Particles
in the darkest region belong to no void over 1 $\sigma$.  The
hatched zones are 1-$\sigma$ subvoids within larger voids.  All zones
are actually subvoids, but most of them do not pass the 1-$\sigma$
level.

\subsection{Selection functions, boundaries, and holes}
\label{realproblems}
Observational effects complicate the application of \zobov\ to real
data, e.g.\ from a galaxy redshift survey.  This section contains some
speculations about how to deal with these effects.

\zobov\ can naturally accommodate a selection function that varies with
position, $\phi(\bm{x})$.  All one needs to do is to divide the
density of each particle/galaxy at ${\bm x}_i$ by $\phi({\bmath x}_i)$
\citep{vdws}.  For void finding, densities estimated with the \dtfe\
(defined between particles) may in general be preferable to what
\zobov\ uses, the \vtfe\ (defined at particles).  However, a variable
selection function is more natural to correct for using the \vtfe.

\zobov\ is designed for a periodic simulation, but other boundary
situations can be handled.  If an isolated set of particles is
analysed without any modifications, \zobov\ will still correctly
determine the adjacencies of each particle.  However, particles on the
edge could have arbitrarily large Voronoi volumes, and thus many
spurious density minima will occur on the edges.  A trivial way of
preventing this is to set all edge particles' densities to a value
higher than any density in the interior.  Another way is to add a
buffer zone of particles at (for example) the mean density around the
dataset.  This will inhibit edge effects for densities estimated for
particles a bit below the surface, as well.  However, there is some
ambiguity in how to make this buffer.

Holes and significantly non-convex boundaries in the data are perhaps
the most difficult problem for \zobov.  A simple way of dealing with
the problem might to put particles in the holes, Poisson-sampling at
the mean density inside them.  The density could also be interpolated
among neighbouring particles, perhaps an iterative process.  Or,
perhaps optimally, it could be estimated through a constrained
realization \citep{bert,hr,zaroubi,vdwbert}.  With any sort of Poisson
hole-filling, it would be wise to try several realizations.

\section{Results}
\label{results}
\aavfcp
\zobovoids 
\probplot

Here I discuss the application of \zobov\ to dark-matter particles
from a cube 60-$\nshMpc$ on a side (hereafter, the `full cube'), taken
from the Millennium simulation.  Aspects of these results in the inner
40-$\nshMpc$ cube (hereafter, the `inner cube') can be found, with
direct comparisons to other void finders, in the Aspen-Amsterdam
Void-Finder Comparison Project (AAVFCP) paper.  To reduce the
dependence on boundary conditions, only voids with cores
(minimum-density particles) in the inner cube were analysed in the
AAVFCP.

\zobov\ is designed for a periodic simulation.  For the non-periodic
AAVFCP cube, a square lattice of particles at the mean density was
added to each face of the full cube, quite far away from the inner
cube (where the results are actually analysed).

\zobov\ detected a couple of orders of magnitude more voids in the
AAVFCP region than almost any other void finder.  This is because of
the many low-significance voids, and subvoids, it detected; see Table
\ref{ratios}, and Figure \ref{density}. The number of 5-$\sigma$ voids
(29) is typical of the number of voids detected by other void finders.

\density

Figures \ref{aavfcp} and \ref{zobovoids} show the largest, and most
significant, \zobov\ void that has a core particle in the inner AAVFCP
cube.  There was actually a deeper void in the full cube, but its
minimum-density particle was on the outer edge, perhaps an artefact of
the boundary conditions used for the tessellation.  It is this void,
not the void shown in the figures, that encompasses nearly the whole
volume, as the most-significant \zobov\ void typically does.

The bottom panel of Fig.\ \ref{zobovoids} shows the void in the top
panel, truncated as in Section \ref{siglevel}, with a 5-$\sigma$
probability threshold.  It is also the most-probable extent of the
void as described in Section \ref{mostsig}, with a caveat.  Figure
\ref{probplot} shows the extent-significance curve for this void,
using two different fits of density-contrast vs.\ probability.  The
solid curve uses the original fit, Eq.\ (\ref{prob}), for $P(r)$ in
Eq.\ (\ref{signif}).  The minimum of this curve (showing the
most-likely extent) is at $i=2$, giving a tiny region around the
central zone.  There are two explanations for this discrepancy between
what \zobov\ and the human eye pick out: density contrast alone is an
inadequate quantifier of void significance (as judged by the human
eye); or, the fit to the probability of void fakeness in Eq.\
(\ref{prob}) is inaccurate at high $r$.

Even using Eq.\ (\ref{prob}), there is a sharp increase in the curve
at $i=200$; the void extent at this point is shown in Fig.\
\ref{zobovoids}.  For this to be returned as the most-probable extent,
the probability of void fakeness must be dramatically reduced by a
factor of e$^{351}$ at $r=4.5$ (the density contrast reached at
$i=200$).  A steeper fit that achieves this is
\begin{equation}
P(r) = \exp[-5.12(r-1)-0.8(r-1)^{4.7}],
\label{probfudged}
\end{equation}
the fit used for the dashed line.  Unfortunately, the Poisson
simulation used for Fig.\ \ref{voidprobdist} is not large enough to
test the probability of such rare, high density contrasts.

\subsection{Lagrangian density distribution}

The top panel of Figure \ref{density} is a scatter plot of density
contrast $r$ versus the core (minimum) density $\rho_{\rm min}$, for
voids in the AAVFCP region.  There are two clusters of points: one for
$\rho_{\rm min}<0.2$, about the natural density of a void in $\LCDM$;
and a second at high density, $\rho_{\rm min} \sim 10^{2.5}$.  At
first, the high-density group may be surprising, but all of these
voids have low density contrasts.  Only one of them barely passes the
3-$\sigma$ level, even fewer than the expected number (10) of
3-$\sigma$ objects in a sample of 3765; this is the number of voids
with $\rho_{\rm min} > 0.2$.  All highly significant voids above $\sim
3$-$\sigma$ in density contrast are also physically significant, with
$\rho_{\rm min}<0.2$.  This lends credence to both significance
measures.

The high-density cluster in the top panel of Fig.\ \ref{density}
appears to be related to the high-density peak at $\rho\approx 10^{2.5}$
in the probability distribution $P(\rho)$ of particle densities, shown
in the bottom panel.  This peak is at approximately the fiducial
density of virialization, $\rho_{\rm vir}\approx 200$, so the
particles in this peak typically reside in collapsed structures.  The
high-density peak in $P(\rho)$ is smaller in the inner cube than in
the full cube, and vice-versa for the low-density peak at $\rho\sim
1$.  This makes sense, since haloes are scarce in the inner cube, most
of which is occupied by a large void.

This $P(\rho)$ is approximately a Lagrangian version of the Eulerian
counts in cells (CIC) statistic $P_{\rm CIC}(N,V)$
\citep[e.g.][]{smn}, which measures the distribution of the
numbers of particles $N$ in fixed grid cells of volume $V$.  Roughly,
$P(\rho) \propto \rho P_{\rm CIC}(N = \rho V,V)$, since, for example,
each cell containing three particles will be counted once for $P_{\rm
CIC}(N=3,V)$, but thrice in $P(\rho=N/V)$.  CIC measurements do not
have a high-density peak, but they often have a significant
high-density tail that, when multiplied by a factor of $\rho$ (or
$N$), may produce a peak.  It would be interesting, but beyond the
scope of this paper, to model this high-density peak in $P(\rho)$
using, for example, the halo model.

\section{Discussion}

\zobov\ has a few unique, appealing features, that I believe are worth
keeping in mind as cosmologists develop a standard definition of
voids.  These features are:

\begin{itemize}
\item Parameter-independence.  The set of voids \zobov\ returns for a
  set of particles depends on no parameters, based on a simple
  definition of a void: a depression around a density minimum.
  However, the word `depression' is also a bit vague; its definition
  for \zobov\ is essentially the first few paragraphs of Section
  \ref{method}.  These implementation choices are, in a sense,
  parameters.

\item Statistical-significance measurement for voids. Void finding is
  not a clear-cut business, so \zobov\ does not return a clear-cut set
  of results.  Instead, it measures a probability that each void is
  real, based on how likely the void's density contrast occurs in a
  Poisson realization.  Thus approach, I believe, is philosophically
  satisfying, but the raw results it returns are not necessarily
  straightforward to analyse.  To get a definitive set of disjoint
  voids, one can set a significance level at which to trust that a
  void is real.  Alternatively, \zobov\ has a mechanism to determine
  the most-probable extents of voids.  There could be ways of
  analysing the raw, parameter-free \zobov\ results, as well.  For
  example, a void probability function measurement could include all
  voids, but weight them by their probability of being real.  \zobov\
  is not the only void-finder that employs a statistical-significance
  test \citep[e.g.][]{kf}.

\item Hierarchical voids.  Just as haloes contain subhaloes, voids
  contain subvoids.  \zobov\ naturally accommodates this fact,
  detecting subvoids as well as voids.  Again, a hierarchy of voids
  does not lend itself to straightforward analysis using traditional
  methods, but methods could be devised which take advantage of this
  hierarchical information.
\end{itemize}

There are also some areas that could benefit from further study or
improvement:

\begin{itemize}
\item Using the (dual) Delaunay instead of the Voronoi tessellation
  for density estimation.  The \dtfe\ and \vtfe\ (based on these two
  tessellations, respectively) give natural density estimates from a
  set of particles \citep{sw,psw,schaap}.  \wvf\ \citep{pwj}, for
  instance, uses the \dtfe\ instead of the \vtfe.  Both the \dtfe\ and
  \vtfe\ have no free parameters, and have infinite spatial
  resolution, up to machine precision.  Arguably, they give the most
  local possible density estimates with meaningful information.  The
  \vtfe\ defines densities at each particle, and thus is natural for
  finding maxima in a set of particles.  This is why we used it for
  the halo finder \voboz.  However, the \dtfe\ is a more natural
  choice for finding minima, since it defines densities in cells
  between particles.  For a well-sampled density field as in an
  $N$-body simulation, the differences are likely negligible, but for
  sparse (e.g.\ galaxy) particle samples, \zobov\ should ideally use
  the \dtfe.  On the other hand, the \vtfe\ could be preferred when
  faced with a variable selection function, which it handles more
  naturally than the \dtfe\ does.

\item The definition of statistical significance for a void.  \zobov\
  judges the statistical significance of a void $v$ by the contrast
  between the lowest density on a ridge beyond which is a deeper void,
  and $v$'s minimum density.  This definition is simple and easy to
  calculate, and the probabilities it returns compare favourably to
  what visual inspection suggests.

  However, there are other possible significance measures.  For
  example, the algorithm could be run several times on different
  Monte-Carlo realizations of the density field, formed by moving
  (`jittering') all particles around in some fashion corresponding to
  the noise in the system.  If the limiting noise is from particle
  discreteness (which is not usually the case for actual data), a
  natural way to jitter the particles would be to move each particle
  to a random place in its initial Voronoi cell.  For $N$-body
  simulations, a jitter according to a measure of the spatial
  resolution (e.g.\ the gravitational softening length) might be more
  appropriate.  For 3D galaxy redshift surveys, the main uncertainties
  are probably the distances inferred from redshifts, and the handling
  of boundary conditions and holes.  For a large data set like an
  $N$-body simulation, this Monte-Carlo approach would probably take
  prohibitively long, but for a more manageable dataset like a galaxy
  catalogue, estimating significances in this way might be tractable.
\end{itemize}

Another finding that emerged from this study is a broad, high-density
peak in the logarithmically binned probability distribution of
dark-matter particle densities, at $\rho \approx 10^3 \bar{\rho}$, as
shown in Fig.\ \ref{density}.  It would be interesting to see whether
this feature can be modelled successfully using the halo model of
large-scale structure.

The code for \zobov, packaged with the halo-finding algorithm \voboz,
is available at \url{http://ifa.hawaii.edu/~neyrinck/voboz}.  The new
edge-detection methods developed for \zobov, described in Section
\ref{edgedetect}, make \voboz\ more attractive than previously for
finding clusters in general point sets, such as galaxy catalogues.

\section*{Acknowledgments}
I thank Istv\'{a}n Szapudi, Erwin Platen, Sergei Shandarin, Andrew
Hamilton, Nick Gnedin and Tom Bethell for helpful discussions, and the
referee, Rien van de Weygaert, for insightful comments and
suggestions.  I also thank Volker Springel for allowing use of
Millennium Simulation dark-matter coordinates, and Rien van de
Weygaert and the Royal Netherlands Academy of Arts and Sciences (KNAW)
for organising the wonderful December 2006 colloquium on Cosmic Voids
in Amsterdam.  I am grateful for support from NASA grant NNG06GE71G
and NSF grant AMS04-0434413.

\end{document}